\documentclass{article}
\usepackage{spconf,amsmath,graphicx,hyperref,amsmath,amssymb,amsfonts,bm}

\usepackage{mathrsfs}
\usepackage{adjustbox}
\usepackage{tikz}
\usepackage{algorithm} 
\usepackage{mathtools}
\usetikzlibrary{positioning}
\usetikzlibrary{quotes,angles}
\usetikzlibrary{calc}
\usetikzlibrary{decorations.pathreplacing}
\usetikzlibrary{shapes.geometric, arrows}
\usetikzlibrary{patterns}
\usepackage{caption}
\usepackage{mathrsfs}
\usepackage{amsfonts}
\usepackage{amssymb}
\usepackage{amsthm}
\usepackage{bbm}
\newtheorem{thm}{Theorem}

\title{Diffusion Denoiser Achievable Analysis for Finite Blocklength Unsourced Random Access}
\twoauthors
 {Yuming Han\sthanks{Corresponding author: Yuming Han.}}
	{Texas A$\&$M University\\
	Department of ECE\\
	College Station, TX 77843, USA}
 {Yuxin Long}
	{University of Leeds\\
	Institute for Transport Studies\\
	Leeds, LS2 9JT, UK}

\begin{document}
%
\maketitle
\begin{abstract}
Polyanskiy proposed a framework for the unsourced multiple access channel (MAC) problem where users employ a common codebook in the finite blocklength regime \cite{Poly_isit_17}. However, existing approaches handle channel noise before the joint decoder. In this work, we introduce a decoder compatible diffusion denoiser as a lightweight analysis within joint decoding. The score network is trained on samples drawn from the channel output distribution, making the method easy to integrate with existing code designs. In our theoretical analysis, we derive a diffusion-denoiser random-coding achievable bound that is strictly tighter. Simulations on existing decoders, including FASURA, MSUG-MRA and pilot-based method, show consistent performance gains with at least a $0.5$ $\mathrm{dB}$ improvement in required $\mathrm{E_b/N_0}$ at a fixed error target.
\end{abstract}
\section{Introduction}
\label{sec:intro}
MAC protocols have recently attracted many research efforts. Specially, a unsourced random access channel (URA) under a finite-blocklength (FBL) constraint was suggested in \cite{Poly_isit_17}. In URA, each user employs the same codebook and the task of the decoder is to recover the list of transmitted messages irrespective of the identity of the users. Recent works  \cite{Poly_trans_20,Fengler_trans_21,Han_allerton, Shyianov_comm_21,Han_tcom} have established fundamental limits for different relevant multiple access channel models. In \cite{kowshik2021fundamental}, the authors studied multi-user interference (MUI) cancellation with and without channel state information. 
\cite{ngo2023unsourced} considered both mis-detection and false alarms, which can serve as a benchmark for unsourced multiple access with random user activities.
However, most recent work analyzes the Gaussian channel while decoding directly from raw outputs. In our work, we add a denoiser at the receiver side to give us a tighter random-coding achievable bound.

Diffusion models gradually adds Gaussian noise to the available training data in the forward diffusion process until the data becomes pure noise. Then, in the reverse sampling process, it learns to recover the data from the noise, which match the denoising process in decoder. In \cite{10104549}, diffusion model is employed to generate the wireless channel for an end-to-end communications system, achieving almost the same performance as the channel-aware case. In \cite{choukroun2023denoising}, diffusion model with an adapted diffusion process is proposed for the decoding of algebraic block codes. \cite{10480348} proposed the channel denoising diffusion models to eliminate the channel noise under Rayleigh fading channel and AWGN channel. To our best knowledge, there is no no theoretical analysis on diffusion model with achievable bound. Most efforts focus on applications and use the denoiser as a pre-processing module at the receiver.

In this work, we embed a diffusion denoiser into the joint decoding analysis, and use it to derive a random-coding achievable bound that is strictly tighter than \cite{Poly_isit_17}. The score network is trained on samples from the channel output distribution so that the approach is highly compatible with existing code designs. Simulations on common decoders (FASURA, MSUG-MRA, pilot-based) confirm at least $0.5$ $dB$ improvements in required 
energy at the same target error rate.

\section{System Model}
\label{sec:system model}
We consider an URA system model from
\cite{Poly_isit_17} with $K_a$ users (or devices). An active user is defined as a user that generates a packet during a given time slot. Each active user generates a packet of $k$ bits. We will assume that an active user does not generate more than one packet at each (discrete) time. Under the power constraint $\|\mathbf{c}\|^2 \leq nP$, we define an $(M,d,\epsilon)$ code, where $M=2^k$ is the size of the codebook, and the error probability $\epsilon \in (0,1)$, consists of:
\begin{itemize}
  \item An encoder $f: [M]\to \mathcal{X}^{n}$ that produces the transmitted codeword $\mathbf{c}_{j}\sim \mathcal N(0, P' I_n)=f(W_j)$ of user $j$ for a given message $W_j$ uniformly distributed over $[M]$.
  \item A decoder $g: \mathcal{Y}^{n}\to  \mathscr{D}([M])$, where $\mathscr{D}([M])$ is a set of indices in $1\ldots M$; the decoder outputs $\{\tilde W_1, \ldots \tilde W_{K_a}\}=g(K_a,\mathbf{y})$ the list of  messages.
\end{itemize}
Let $\tilde g(K_a,\mathbf{y})$ be the multiset of messages. Then we define the $j$-th error event as follows

\begin{equation}
    \begin{aligned}
  E(W_j)=\{&W_j\in \tilde g(K_a,\mathbf{y})\cap W_j\notin g(K_a,\mathbf{y}),\\
  &\text{ or $W_j$ repeated in $\tilde g(K_a,\mathbf{y})$}\}.
  \label{eq:error event}
\end{aligned}
\end{equation}

As in \cite{Poly_isit_17} there are two type of errors: if, for a message $W_j$ transmitted by some user, $W_j\notin g(K_a,\mathbf{y})$ an error happens. If two or more users transmit the same message with the same decoding time, a collision  happens, which may be counted as an error. However, as in \cite{Poly_isit_17} the probability of collision for realistic parameters is extremely small and does not influence performance.
We require that the steady state error is bounded as $ \frac{1}{K_a}\sum_{W_j\in \tilde g(K_a,\mathbf{y})} \mathbb{P}(E(W_j))\leq\epsilon.$
Notice that we do not require the
error probability at each active user to be
bounded by $\epsilon$, only the all users' averaged error
probability.

In the remainder of this paper we exclusively focus
on the Gaussian MAC (GMAC). The received signal is given by
\begin{equation}
\mathbf{y} \;=\; \sum_{i=1}^{K_a} \mathbf{c}_i \;+\; \mathbf{Z}, 
\qquad \mathbf{Z} \sim \mathcal N(0, \mathbf{I}_n).
\label{eq:channel}
\end{equation}
Following \cite{Poly_isit_17} we only care about obtaining a list of messages, not from which transmitters they are transmitted. In our case, the list $g(K_a,\mathbf{y})$ consisting of estimates
of messages transmitted with $K_a$ active users. It should be noted that the receiver is assumed to know how many messages are generated
at each point in time, but not which users generate them.
Without this knowledge, the
receiver would have to estimate the number of active
users $K_a$ at each point of time. This has been
considered in a number of papers \cite{ngo2023unsourced,tensor-based-estimate-ka, many-user-estimate-ka, pilot-based-estimate-ka}.

\section{Diffusion Denoiser Achievable Bound}
\label{sec:method}
\subsection{Achievable Bound}
We propose a diffusion denoiser model at the receiver within joint decoding. We give a random-coding achievable bound below. This bound is first established by upper-bounding the probability of $t$-misdecoded messages, followed by an information density analysis to further refine the error probability.
\begin{thm}
(Diffusion Denoiser Random-coding Bound): Fix $P^{\prime} < P$, there exists an $(M, d, \epsilon)$ random-access code with constants $J_* \geq 0$ and $K_E \geq 0$ satisfying the power constraint $P$ and
\begin{align}
    \epsilon \leq \sum_{t=1}^{K_a}\frac{t}{K_a}\min(q_1(t),q_2(t))+q_0,
    \label{Thm2}
\end{align}
where
\begin{align}
    q_0=\frac{1}{M}\binom{K_{a}}{2}+ K_{a}\frac{\Gamma(n,nP/P^{\prime})}{\Gamma(n)},
\end{align}
\begin{align}
q_1(t) =\mathop{\max}\limits_{\rho_2,\rho_3 \in [0,1]} &\binom{K_a}{t}^{\rho_1}\binom{M-K}{t}^{\rho_0\rho_1} \cdot \notag\\
&\exp \left( - \frac{\rho_0\rho_1}{2tP'+16v_*} \right)\Big(1+\tfrac{tP'}{8v_*}\Big)^{-n\rho_0\rho_1},
\end{align}
\begin{equation}
q_2(t)=\inf_{\psi}\left\{\mathbb{P}[I_t\leq \psi]+\exp\left\{n(tR_1+R_2)/2-\psi\right\}\right\},
\end{equation}
\begin{align}
I_t=\min_{S_0\subset [K_a],|S_0|=t} &\frac{n}{2}\log(1+P^{\prime}t)+\frac{||\mathbf{y}-c(S_0^c)||^2_2}{1+P^{\prime}t}\notag\\
&-||\mathbf{y}-c(S_0)-c(S_0^c)||^2_2,
\end{align}
where $v_*=1 + J_* + K_E^2$, $R_1=\frac{1}{nt} \log \binom{M-K_a}{t}$, $R_2=\frac{1}{n} \log \binom{K_a}{t}$, $\mathbf{y} \sim \mathcal{N}(\mathbf{0},(1+K_aP^{\prime})\mathbf{I}_{n})$ and $\Gamma(x,y)=\int_y^\infty z^{x-1}e^{-z} dz$.
\label{thm_BT}
\end{thm}
As in \cite{Poly_isit_17}, we analyze the probability of error after we replace the measure over which the expectation is taken by the one under which: i) all active users transmit distinct messages; 
ii) each user transmits $\mathbf{c}_i$ instead of $\mathbf{c}_i \mathbbm{1} \{ ||\mathbf{c}_i||_2^2 \leq nP\}$ under the true measure. The total variation $d_{TV}$ between the original measure and the new one can be upper-bound by:
\begin{equation}
    \begin{aligned}
d_{TV}&=\mathbb{P}[\text{$W_j$ repeated}]+\mathbb{P}[||\mathbf{c}_i||^2 > nP, \forall i\in K_a]\\
&\leq \frac{1}{M}\binom{K_{a}}{2}+ K_{a}\frac{\Gamma(n,nP/P^{\prime})}{\Gamma(n)}=q_0,
\label{Gamma}
    \end{aligned}
\end{equation}
where (\ref{Gamma}) holds since $||\mathbf{c}_i||^2$ follows the Gamma distribution with shape $n$ and scale $P^{\prime}$. Now after the change of measure, the probability of error can be bounded as
\begin{align}
\epsilon \leq \sum_{t=1}^{K_a}\frac{t}{K_a}\mathbb{P}[t\text{-misdecoded}]+q_0.
\label{epsilon_BT}
\end{align}
We then proceed to bound $\mathbb{P}[t\text{-misdecoded}]$ using two distinct methods. The first approach stems from the encoder's perspective, yielding the term $q_1(t)$ in (\ref{eq:q1}). Meanwhile, the second method adopts the decoder's viewpoint, giving rise to the term $q_2(t)$ in (\ref{eq:q2}).

First, we define the sum-codewords as $c(S)=\sum_{i\in S}\mathbf{c}_i$. We assume $t$ of $K_a$ messages are mis-decoded, which is the same event as when we let $S_0 \subset [K_a]$ of messages be replaced with $S_0^{\prime} \subset \{K_a+1,\cdots,M\}$ and $|S_0|=|S_0^{\prime}|=t$. Then let $F(S_0,S_0^{\prime})$ denote the set of $(S_0,S_0^{\prime})$ such that $\{||c(S_0)-c(S_0^{\prime})+\mathbf{Z}|| < ||\mathbf{Z}|| \}$ holds. Now, we define the residual of our diffusion denoiser $D$ as
\begin{align}
    \mathbf{R}\triangleq D(\mathbf{y}) \;-\; \sum_{i=1}^{K_a} \mathbf{c_i}.
\end{align}
We set $s(\mathbf{y}) \;\triangleq\; \nabla \log p_Y(\mathbf{y})$ be the score function of recieved signal. The residual can be decomposed as $\mathbf{R} = \mathbf{Z} + s(\mathbf{y}) + d(\mathbf{y}),$ where $d(\mathbf{y}) \triangleq\; D(\mathbf{y}) - \big(\mathbf{y} + s(\mathbf{y})\big)$ is a mismatch term. Then, we rewrite the  pairwise error event
\begin{align}
    F(S_0, S_0^{\prime}) &= \{||c(S_0)-c(S_0^{\prime})+\mathbf{R}|| < ||\mathbf{R}||\} \notag \\
    &=\{2 h^T\mathbf{R}+||h||^2 \},
\end{align}
where $h=c(S_0)-c(S_0^{\prime})$.
We rewrite the error event as a union of $F(S_0,S_0^{\prime})$ to have
\begin{align}
\mathbb{P}[t\text{-misdecoded}] \leq \mathbb{P}\left[\bigcup_{S_0\in \binom{K_a}{t}} \bigcup_{S_0^{\prime}\in \binom{M-K_a}{t}} F(S_0,S_0^{\prime})\right].
\end{align}
By exponential Markov’s inequality, for any $\gamma>0$, we have
\begin{align}
\mathbb P\!&\left[F(S_0,S_0^{\prime}) \,\big|\, h, \mathbf{Z}\right]
\leq e^{-\gamma \|h\|^2}\, \notag\\
&\big(\mathbb E [e^{-4\gamma h^\top \mathbf{Z}}\big])^{1/2}
\big(\mathbb E [e^{-4\gamma h^\top s(\mathbf{y})}\big])^{1/2}
\big(\mathbb E [e^{-4\gamma h^\top d(\mathbf{y})}\big])^{1/2}.
\end{align}
BY direct computing with moment-generating function, there exist finite constants $J_* \geq 0$ and $K_E \geq 0$
such that
\begin{align}
\mathbb P\!\left(F(S_0,S_0^{\prime}) \,\big|\, h, \mathbf{Z}\right)
\leq \exp\!\Big(-\gamma\|h\|^2 + 4\gamma^2 v_* \|h\|^2\Big),
\end{align}
where $v_* \;=\; 1 + J_* + K_E^2$. Optimizing over $\gamma$ gives $\gamma^* = 1/(8 v_*)$ and the pairwise bound would be
\begin{equation}
\mathbb P\!\left(F(S_0,S_0^{\prime}) \,\big|\, h,\mathbf{Z}\right)
\;\leq\;
\exp\!\left(-\frac{\|h\|^2}{16 v_*}\right).
\label{eq:pairwise}
\end{equation}
Since $h\,|\,t \sim \mathcal N(0, 2tP' I_n)$, we have
\begin{equation}
\mathbb P\!\left(F(S_0,S_0^{\prime}) \,\big|\, h, \mathbf{Z}\right)
\;\leq\;
\exp \left( - \frac{1}{2tP'+16v_*} \right)\Big(1+\frac{tP'}{8v_*}\Big)^{-n} .
\label{eq:ball}
\end{equation}
Using Gallager’s $\rho$-trick for any $\rho_0 \in [0,1], \rho_1 \in [0,1]$ to get
\begin{align}
\mathbb P[t&\text{-misdecoded}] \leq \binom{K_a}{t}^{\rho_1}\binom{M-K_a}{t}^{\rho_0\rho_1} \cdot \notag\\
&\exp \left( - \frac{\rho_0\rho_1}{2tP'+16v_*} \right)\Big(1+\tfrac{tP'}{8v_*}\Big)^{-n\rho_0\rho_1}  = q_1(t)
\label{eq:q1}
\end{align}
Second, we introduce an alternative bound for $\mathbb{P}[t\text{-misdecoded}]$. The information density can be defined as 
\begin{align}
i_t(c(S_0);\mathbf{y}|c(S_0^c))=&\frac{1}{n}\log(1+P^{\prime}t)+\frac{||\mathbf{y}-c(S_0^c)||^2_2}{1+P^{\prime}t}\notag\\
&-||\mathbf{y}-c(S_0)-c(S_0^c)||^2_2.
\label{eq:infor density}
\end{align}
Note that $F(S_0,S_0^{\prime})$ is equivalent to $\{i_t(c(S_0^{\prime});\mathbf{y}|c(S_0^c)) > i_t(c(S_0);\mathbf{y}|c(S_0^c))\}$.
Let $I_t=\min_{S_0} i_t(c(S_0);\mathbf{y}|c(S_0^c))$. For a fixed $\psi$, we have
\begin{equation}
    \begin{aligned}
\mathbb{P}[t\text{-misdecoded}] &\leq \inf_{\psi} (\mathbb{P}[I_t \leq \psi]\\
&+\binom{K_a}{t} \binom{M-K_a}{t}e^{-\psi})=q_2(t).
\label{eq:q2}
\end{aligned}
\end{equation}
This completes the proof of Theorem 1. Now we have a achievable bound for random-coding with constants $J_*$ and $K_E$. In practice, we need to train a score-diffusion network $s_{\theta}(\mathbf{y})$. The denoiser samples the trained score network to give an optimization of $J_*$ and $K_E$ as follows.
\subsection{Diffusion Model}
Diffusion models define a generative framework by coupling a \emph{forward diffusion process} with a learned \emph{reverse denoising process}.
In discrete time, the forward process gradually perturbs a clean data point $x_0 \sim p_{\text{data}}$ using a predefined variance schedule $\{\beta_l\}_{l=1}^T$:
\begin{equation}
    q(x_l \mid x_{l-1}) = \mathcal{N}\!\left(\sqrt{1-\beta_l}\,x_{l-1},\, \beta_l I\right),
    \quad l=1,\dots,T,
\end{equation}
which admits the closed form
\begin{equation}
    q(x_l \mid x_0) = \mathcal{N}\!\left(\sqrt{\bar{\alpha}_l}\,x_0,\,(1-\bar{\alpha}_t)I\right),
    \quad \bar{\alpha}_l = \prod_{s=1}^l (1-\beta_s).
\end{equation}
As $l \to T$, the distribution approaches an isotropic Gaussian, effectively destroying the input signal.  An equivalent continuous-time formulation expresses the forward dynamics as a stochastic differential equation (SDE):
\begin{equation}
    dx_l = f(x_l, l)\,dl + u(t)\,dw_l, \quad l \in [0,T],
\end{equation}
where $w_l$ is a standard Wiener process. The corresponding \emph{reverse-time SDE} is given by
\begin{equation}
    dx_l = \Big[f(x_l, l) - u(t)^2 \nabla_{x_l}\log q_l(x_l)\Big]\,dl + u(l)\,d\bar{w}_l,
\end{equation}
where $l \in [T,0]$ and $\bar{w}_l$ denotes a reverse-time Wiener process, and $\nabla_{x_l}\log q_l(x_l)$ is the score function guiding the generative dynamics.  
Since the true score is intractable, it is approximated by a neural network $s_\theta(x_l,l)$ trained via denoising score matching:
\begin{align}
    \mathcal{L}_{\text{score}}(\theta) =
    \mathbb{E}_{t \sim \mathcal{U}[0,T]} \,
    &\mathbb{E}_{x_0 \sim p_{\text{data}}, \, x_t \sim q_t(x_t|x_0)}
    \Big[\lambda(t)\\ \notag
    &\,\|s_\theta(x_t,t) - \nabla_{x_t}\log q_t(x_t|x_0)\|^2 \Big],
\end{align}
where $\lambda(l)$ is a time-dependent weighting term.  
In practice, this objective reduces to predicting the Gaussian noise that was added during the forward process. Writing $ x_l = \sqrt{\bar{\alpha}_l}\,x_0 + \sqrt{1-\bar{\alpha}_l}\,\epsilon,\epsilon \sim \mathcal{N}(0,I),$
as the loss becomes the simplified noise-prediction objective:
\begin{equation}
    \mathcal{L}_{\text{noise}}(\theta) =
    \mathbb{E}_{x_0, \epsilon, l}\!
    \left[\big\|\epsilon - \epsilon_\theta(x_l, l)\big\|^2 \right].
\end{equation}
Thus, training a diffusion model amounts to teaching the network to recover the injected noise $\epsilon$ from a corrupted sample $(x_l,l)$. During inference, sampling begins with Gaussian noise $x_T \sim \mathcal{N}(0,I)$, and the learned reverse dynamics are applied iteratively to denoise $l_T \to l_0$. Equivalent ODE-based formulations provide both stochastic and deterministic generation procedures.

In our model, we set the observation with the diffusion state at level $l^*$ with $s_{\theta}(\mathbf{y})\triangleq s_{\theta}(\mathbf{y},l^*)$. $J_*$ and $K_E$ characterize the concentration of the score function and the variance of the denoiser mismatch, respectively. Both parameters are estimated empirically from finite samples of the score network outputs and residuals.
To approximate the spectral bound of the score covariance, we first sample $N$ of the channel outputs $\mathbf{y}$ to have $\mathbf{y}^{(k)}$. For each sample, the trained score network produces $s_{\theta}^{(k)}=s_{\theta}(\mathbf{y}^{(k)})$. The empirical second moment is then given by $\hat{J}=\frac{1}{N}\sum_{k=1}^N s_{\theta}^{(k)} s_{\theta}^{(k)T}$. Then, we extract the maximum eigenvalue via power iteration as $J_*=\lambda_{\max}(\hat{J}).$
This procedure ensures that the largest variance direction of the score distribution is captured reliably.

For the denoiser residual $K_E$, we employ a similar process. We can compute the mismatch residual using $s_{\theta}(\mathbf{y})$ as $E_{\theta}(\mathbf{y})=D(\mathbf{y})-(\mathbf{y}+s_{\theta}(\mathbf{y})).$
By sampling $\mathbf{y}^{(k)} \sim \mathbf{y}$, we calculate the residuals $r^{(k)}=E_{\theta}(\mathbf{y}^{(k)})$. Then, we can have the empirical covariance matrix as $K_E=\frac{1}{N}\sum_{k=1}^N \tilde{r}^{(k)} \tilde{r}^{(k)T}$,
where $\tilde{r}=r^{(k)}-\frac{1}{N}\sum_kr^{(k)}$. Here the top eigenvalue of $K_E$ provides a conservative estimate of the residual’s directional sub-Gaussian parameter. 
Our $J_*$ and $K_E$ require only forward evaluation of the trained score network and the denoiser. These constants can then be directly integrated into the finite-blocklength bound.
\section{SIMULATION RESULTS}
\begin{figure}
    \centering
    \includegraphics[height=7cm,width=8cm]{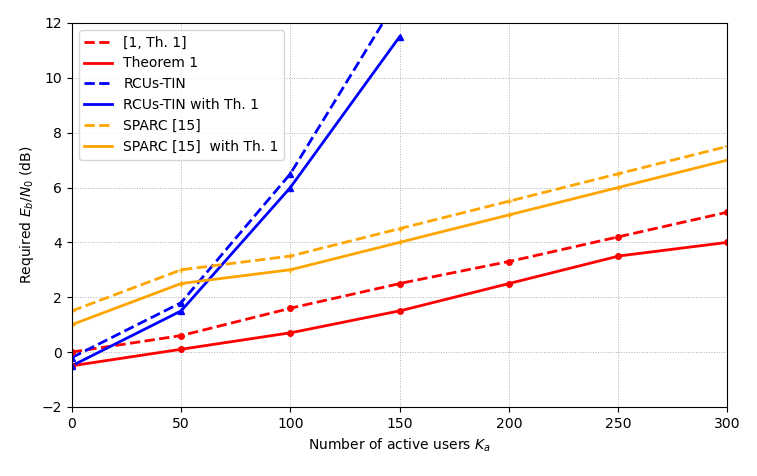}
    \caption{The required energy per bit to achieve $\epsilon=0.001$ as a function of the number of active users for Theorem 1 and different achievable schemes.}
    \label{fig:1}
\end{figure}
First, we plot numerical results for the bounds developed above in Figure. \ref{fig:1}. We compare various strategies in the following settings. Each active users is  transmitting $k=100$ bits of information. The frame length is $n=30000$ and the target per-user probability of error is $0.001$. We first compare our achievable bound against the classical result of \cite{Poly_isit_17}. Our bound reduces the required energy by approximately $0.8$ $dB$. We then instantiate Theorem~1 for two representative schemes, RCU-TIN and SPARC \cite{sparcs}. In each case, the required energy drops by about $0.5$ $dB$ at the same error target.

\begin{figure}
    \centering
    \includegraphics[height=7cm,width=8cm]{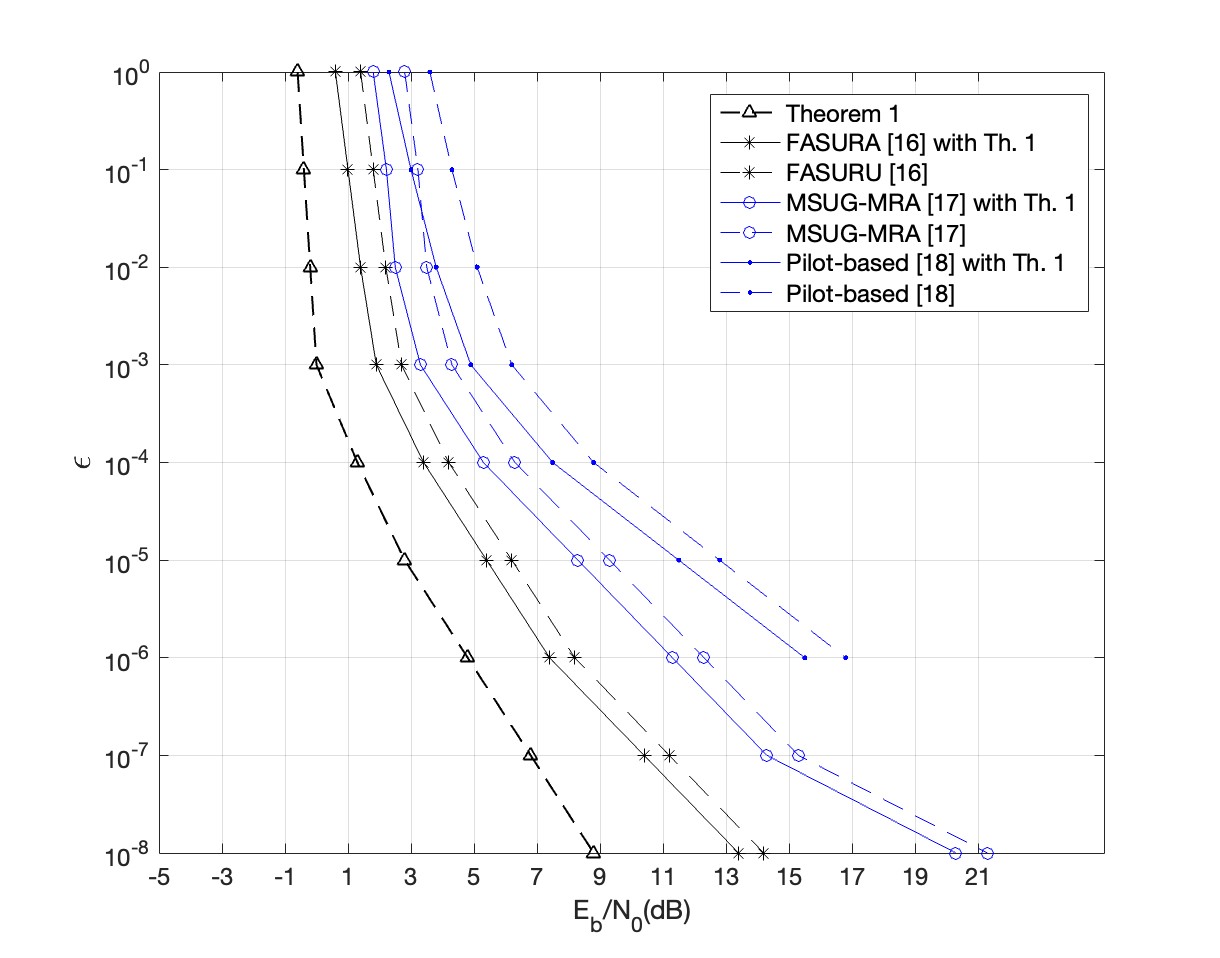}
    \caption{The bounds on the probability of error $\epsilon$ as a function of \( E_b/N_0 \) for Theorem 1 and different achievable schemes.}
    \label{fig:2}
\end{figure}
Second, we compare our approach with \cite{FASURA,msug,piolt} across a range of target error probabilities in Figure. \ref{fig:2}. For each decoder, augmenting with Theorem~1 reduces the required energy to meet a given error target by approximately $0.5$ $dB$ relative to the original scheme. This consistent gain demonstrates that our theorem is compatible and effective across state-of-the-art coding designs.
\section{CONCLUSIONS}
\label{sec:foot}
We presented a decoder compatible diffusion denoiser that is integrated into the joint decoding for URA in the finite-blocklength regime. First, we derived a diffusion denoiser achievable bound that is strictly tighter than \cite{Poly_isit_17}. The score network is trained on channel output samples, enabling minimal-friction use with existing code designs. Empirically, we employ our theorem in decoders, such as FASURA, MSUG-MRA, and pilot-based scheme. Our approach consistently reduces the required energy by at least $0.5$ $dB$ at a fixed error target.

\bibliographystyle{IEEEbib}
\bibliography{strings,refs}

\end{document}